\documentclass[10pt]{article}
\usepackage{a4wide}                      
\usepackage{epsf}                        
\usepackage{latexsym}                    
\usepackage{amsfonts}                    
\usepackage{amssymb}                     

\newcommand{\sect}[1]{\setcounter{equation}{0}\section{#1}}
\renewcommand{\theequation}{\arabic{section}.\arabic{equation}}

\def\be{\begin{equation}}
\def\ee{\end{equation}}
\def\bea{\begin{eqnarray}}
\def\eea{\end{eqnarray}}
\def\nn{\nonumber \\[.2cm]}
\def\vsp#1{\vspace{#1}}
\def\hsp#1{\hspace{#1}}
\def\Ortin{Ort{\'\i}n}
\def\iac{\'\i}

\def\part{\partial}
\def\tfrac#1#2{{\textstyle{\frac{#1}{#2}}}}
\def\half{\tfrac{1}{2}}

\def\Str{\mbox{STr}}

\def\incl{\mbox{i}}
\def\cL{{\cal L}}  
\def\ux{{\underline x}}
\def\hV{\hat V} 
\def\hD{\hat D} 
\def\hF{\hat F}  
\def\ha{{\hat a}}     
\def\hb{{\hat b}}     
\def\hzeta{{\hat{\zeta}}}  
\def\hmu{{\hat{\mu}}}                  
\def\hnu{{\hat{\nu}}}    
\def\hrho{{\hat{\rho}}}    
\def\hchi{{\hat{\chi}}}

\def\cF{{\cal F}}  

\def\mn{{\mu\nu}}

\def\hmn{{\hmu\hnu}}

\def\makeatletter{\catcode`\@=11}
\makeatletter
\def\mathbox#1{\hbox{$\m@th#1$}}%
\def\math@ccstyles#1#2#3#4#5#6#7{{\leavevmode
      \setbox0\mathbox{#6#7}%
      \setbox2\mathbox{#4#5}%
      \dimen@ #3%
      \baselineskip\z@\lineskiplimit#1\lineskip\z@
      \vbox{\ialign{##\crcr
             \hfil \kern #2\box2 \hfil\crcr
             \noalign{\kern\dimen@}%
            \hfil\box0\hfil\crcr}}}}
\def\mathaccstyles{\math@ccstyles\maxdimen}
\def\maththroughstyles{\math@ccstyles{-\maxdimen}}
\def\unity%
 {\maththroughstyles{.45\ht0}\z@\displaystyle {\mathchar"006C}\displaystyle 1}
\begin{document}

\rightline{KUL-TF/05-18}
\rightline{UG-FT-192/05}
\rightline{CAFPE-62/05}
\rightline{hep-th/0507198}
\rightline{July 2005}
\vspace{1truecm}

\centerline{\LARGE \bf On the gauge invariance and coordinate transformations}
\vspace{.6truecm}
\centerline{\LARGE \bf  of non-Abelian D-brane actions}
\vspace{1.3truecm}

\centerline{
    {\large \bf Joke Adam${}^{a,}$}\footnote{E-mail address: 
                                  {\tt joke.adam@fys.kuleuven.ac.be}},
    {\large \bf Ignacio A. Ill\'an${}^{b,}$}\footnote{E-mail address: 
                                  {\tt ignacio@physik.hu-berlin.de}}
    {\bf and}
    {\large \bf Bert Janssen${}^{b,}$}\footnote{E-mail address: 
                                  {\tt bjanssen@ugr.es} }
                                                            }
\vspace{.4cm}
\centerline{{\it ${}^a$ Instituut voor Theoretische Fysica, K.U. Leuven,}}
\centerline{{\it Celestijnenlaan 200D,  B-3001 Leuven, Belgium}}

\vspace{.4cm}
\centerline{{\it ${}^b$ Departamento de F\'{\i}sica Te\'orica y del Cosmos and}}
\centerline{{\it Centro Andaluz de F\'{\i}sica de Part\'{\i}culas Elementales,}}
\centerline{{\it Universidad de Granada, 18071 Granada, Spain}}

\vspace{2truecm}

\centerline{\bf ABSTRACT}
\vspace{.5truecm}

\noindent
We study the variations of the worldvolume fields in the non-Abelian action for multiple 
D-branes. Using T-duality we find that the embedding scalars transform non-trivially under 
NS-NS gauge transformations as $\delta X \sim [X, X]$ and prove that the non-Abelian 
Chern-Simons action is invariant under these transformations. Given that T-duality relates 
the (part of the) NS-NS transformation with (part of the) general coordinate transformations, 
we can get some insight in the structure of non-Abelian coordinate transformations.

\newpage
\sect{Introduction}

The dielectric effect \cite{Myers} is by now a well established phenomenon in modern 
string theory. The effect is a manifestation of the non-Abelian physics of multiple 
coinciding D-branes, which is a direct consequence of the $U(1)^N \rightarrow U(N)$ gauge 
enhancement in the worldvolume theory of a stack of D-branes \cite{Witten}. Applications of the 
dielectric effect can be found in numerous parts of string theory: the polarisation of 
D$p$-branes into a single fuzzy D$(p+2)$-brane \cite{Emparan, Myers, TV}, gravity duals 
of confining gauge theories \cite{PS, KS}, enhan\c{c}ons \cite{Johnson}, matrix models in 
non-trivial backgrounds \cite{BMN}, a microscopic description of giant gravitons 
\cite{DTV, MT, JL} and attempts to construct longitudinal 5-branes in Matrix theory 
\cite{HP, LR}, just to mention a few.

The effect is due to a number of non-Abelian terms in the Chern-Simons part of the worldvolume 
action of $N$ coinciding D$p$-branes, which allow the D$p$-branes to couple to R-R forms of
rank $n > (p+1)$. The presence of these non-Abelian couplings can be derived in several ways: 
either via Matrix theory methods \cite{TvR}, at least for weakly curved backgrounds, or via 
T-duality, requiring that the Chern-Simons action is of the same form as the duality maps a 
D$p$ into a D$(p-1)$-brane~\cite{Myers}, generalising to the non-Abelian case the calculation 
of \cite{ABB, BR}. It was shown that in this case the $U(N)$ covariant derivative 
$DX$ of the pullback, necessary in order for the action to transform well under $U(N)$ gauge 
transformations \cite{Dorn, Hull}, is mapped by T-duality into a commutator term $[X, X]$. 
These commutator terms are usually referred to as the dielectric terms, as they allow solutions 
of polarised brane configurations.

Although the dielectric terms arise naturally form Matrix theory and T-duality and are 
responsible for numerous successes in string theory, the issue of the gauge invariance of the 
Chern-Simons action including these dielectric terms has so far mostly been neglected. In 
\cite{GHT} a  first attempt was made to clarify the gauge invariance of non-Abelian D-brane 
actions, however this was done before the work of \cite{Dorn, Hull, Myers}, so neither the 
covariant pullback, nor the dielectric terms were taken into account. The problem of the 
gauge invariance of the dielectric terms was first tackled in \cite{Ciocarlie} for the R-R 
transformations $\delta C_p = \partial \Lambda_{p-1}$ by writing the action as a series of 
terms that involve only the R-R field strengths. The results are quite limited though, as they 
do not take into account the full R-R transformations $\delta C_p = \sum_k \partial 
\Lambda_{p-2k-1}B^k$, the NS-NS transformation $\delta B = \partial \Sigma$  and the 
massive gauge transformation $\delta C_p = -m  \Sigma B^{(p-1)/2}$. 

A different approach was taken in \cite{AGJL}. Here, instead of writing 
the action in an explicitly invariant way, the variations of the background fields in the 
non-Abelian action were reviewed. It was shown that, in order for the action to transform at 
the same time as a total derivative, as a scalar under $U(N)$ and with the correct Abelian 
limit, the transformation of pullback of the R-R fields in the non-Abelian action had to be 
defined as\footnote{We use the shorthand notation $(\incl_X \incl_X) C_p$ for the contraction 
of the commutators with the bulk fields, defined as
\[
(\incl_X \incl_X C)_{\mu_1 ... \mu_n} = \half [X^\rho, X^\nu] C_{\nu\rho\mu_1 ... \mu_n}.
\]  }
\be
\delta P [(\incl_X \incl_X)^r C_p] = D P[(\incl_X \incl_X)^r \Lambda_{p-1}] 
\label{defC}
\ee
for the variation $\delta C_p = \partial \Lambda_{p-1}$ (for the definition of the full R-R 
transformation rules involving the Kalb-Ramond field $B$, we refer to \cite{AGJL}). 

It was also shown that the action could be written in such a way that the Kalb-Ramond field 
either combines with the Born-Infeld field strength $F=2 \partial V + i [V, V]$ into the 
combination $\cF = F + P[B]$ or appears contracted with commutator terms in the form 
$(\incl_X \incl_X B)$ and $(\incl_X B)$. Defining the variation of the pullback of $B$ 
analogously to (\ref{defC})
\be
\delta P [B] = 2 D P[\Sigma], 
\ee
it was argued that $\cF$ was gauge invariant, while the variation of the remaining terms 
$(\incl_X \incl_X B)$ and $(\incl_X B)$ was zero, due to the isometries in the transverse 
space necessary in order to perform the T-dualities. The presence of the isometries was 
believed to be related to the problem of how to perform general coordinate transformations 
in a non-commutative geometry. 

There are reasons to believe, however, that these arguments are not quite true. A first hint 
comes from the fact that for the dielectric effect, the expansion of $N$ D$p$-branes into a 
spherical, fuzzy D$(p+2)$-brane, the dependence on the transverse direction is necessary. 
In some way, the isometry direction can be removed and the dependence restored after the 
T-duality is performed, just in the same way as in the Abelian case.

This leaves us with two questions, an immediate and a deeper one. The immediate question is: 
what happens to the NS-NS gauge invariance of the Chern-Simons action and more precisely, what 
is the variation of the contracted Kalb-Ramond fields $(\incl_X \incl_X B)$ and $(\incl_X B)$? 
Do we need to add more terms to the action in order to cancel these variations or is the action 
invariant in its present form, as derived from T-duality in \cite{Myers}? We will show in this 
letter that all the fields in the action (both target space as worldvolume) transform 
non-trivially under NS-NS transformations and that the action is in fact invariant.

The deeper question is about the role played by the non-commutative general coordinate 
transformations in a non-Abelian action. The restoring of the dependence of transverse coordinates 
after T-duality is not the only example of non-commutative diffeomorphisms playing a subtle role 
in the calculation. In \cite{JLR, LR} an effective action for non-Abelian gravitational waves 
was used, which strictly speaking is only valid after the coordinate transformation 
$X \rightarrow Y$ and $Y\rightarrow -X$. Although it is not clear how to perform such a 
coordinate transformation in a rigorous way, the results reproduce precisely the known 
results of the Abelian limit.

There have been numerous studies about non-commutative diffeomorphism invariance in string 
theory \cite{DS, Hassan, DSW, BFLV, CK}, ranging from worldsheet arguments to geometrical 
approaches. In this paper we would like to focus the issue from a different perspective. 
As we are originally interested in the gauge invariance of the non-Abelian Chern-Simons 
action, we will try to obtain some information about the structure of coordinate transformations 
from these gauge transformations. 

It is well known that T-duality interchanges winding and momentum modes, or in other words 
certain components of the metric with certain components of the Kalb-Ramond field. It is 
then clear that the variations of the Kalb-Ramond field (i.e. NS-NS gauge transformations) 
are mapped by T-duality into variations of the metric (i.e. coordinate transformations), 
at least part of each of these. We will apply  T-duality on the variations of the worldvolume 
fields and derive their transformation rules.

The paper is organised as follows: in section 2 we derive carefully the variations 
of the worldvolume fields (embedding scalars and Born-Infeld vector) under worldvolume and 
target space diffeomorphisms, $U(N)$ gauge transformations and NS-NS gauge transformations, 
as implied by T-duality. In particular, we will see that the non-Abelian embedding scalars 
transform non-trivially under NS-NS transformations. In section 3 we concentrate on this 
NS-NS variation and derive the precise form of the variations of the different constituents 
of the action and in section 4 we prove the invariance of the non-Abelian D6-brane Chern-Simons 
action under these transformations. The general case of the non-Abelian D$p$-brane action 
is left for Appendix A.

\sect{Variation of the worldvolume fields through T-duality}

The T-duality between a D$p$ and a D$(p-1)$-brane implies, at the level of the effective 
actions, that the physical content of both effective actions is equivalent. In particular, 
there exists a one to one map between the field content of both actions. Furthermore, the 
role played by specific fields in one action should be the same as the role played by these 
fields in the other one. It turns out that from these two properties we can derive 
information about the form of the action and the variation of the worldvolume fields. 

Let us first look briefly at the Abelian case, in order to set our conventions. A well known 
consequence of the first property is the mixing of the embedding scalars $X$ and the Born-Infeld 
(BI) vector $V$ under T-duality \cite{BR}: whereas the bosonic field content of the D$p$-brane 
consists of a $(p+1)$-dimensional $U(1)$ vector $\hat V_{\hat a}$ and $(9-p)$ transverse scalars 
$Y^i$, the field content of the D$(p-1)$-brane contains a vector $V_a$ in one dimension less, 
but has one extra transverse scalar $X^{\ux}$. The T-duality between the  D$(p-1)$-brane 
and the D$p$-brane, performed in one of the worldvolume directions of the D$p$-brane, therefore 
maps the extra component of $\hat V_{\hat a}$ into the extra transverse scalar of the 
$X^{\hat \imath}$'s. Concretely, the T-duality map between the bosonic worldvolume fields is 
given by\footnote{Our notation for this section is as follows: we split the worldvolume indices 
$\hat a = (a, x)$ and the transverse indices $\hat \imath = (i, \ux)$, where  we denote by 
$\sigma^x$ the worldvolume direction of the D$p$-brane in which the T-duality is performed 
and by $Y^\ux$ the target space direction corresponding to the same. In the same way we split 
the full target space indices $\hmu=0, ..., 9$ into $\mu= 0, ..., 8$ and $\ux$.}
\bea
&& \hat V_a \longrightarrow V_a, 
\hsp{3cm}
Y^i \longrightarrow X^i, \nonumber \\ [.2cm]
&& \hat V_x \longrightarrow X^\ux,
\hsp{2.85cm}
Y^{\ux} \ = \ \sigma^x.
\label{TdualWV}
\eea 
The last equation is merely an expression of the fact that we write the actions in the static 
gauge, at least the direction in which the T-duality is performed, while the first two state 
that the BI vector components and the transverse scalars in directions different from the 
T-dualised one are the same in both actions. The third equation is the one that matches the
 bosonic degrees of freedom, mapping the $x$-component of the D$p$-brane BI vector with the 
extra embedding scalar of the D$(p-1)$-brane.   

It is easy to show that under the above transformations (\ref{TdualWV}) the BI field 
strength $\hat F_{\hat a \hat b} = 2 \part_{[\hat a} V_{\hat b]}$ and the pull-back 
$\part_{\hat a} Y^i$ transform under T-duality as:
\bea
&&
\part_a Y^i \longrightarrow \part_a X^i, \hsp{1cm}
\part_x Y^i = 0, \hsp{1cm}
\part_{\hat a} Y^\ux = \delta^\ux_{\hat a}, \nn
&&
\hsp{1cm}
\hat F_{ax} \longrightarrow \part_a X^\ux, \hsp{1cm}
\hat F_{ab}  \longrightarrow F_{ab}.
\label{Tdualfieldstr}
\eea
These duality rules, together with the duality rules for the background fields \cite{BHO}, 
guarantee the fact that the action of the D$p$-brane is mapped into the action of the 
D$(p-1)$-brane \cite{ABB, BR}. 

The second conclusion, namely that the role played by each field in both actions should be the 
same, implies that the variations of the worldvolume fields should have the same form in both 
actions. In particular, the BI vector $\hV_\ha$ of the D$p$-brane transforms as a vector under 
worldvolume coordinate transformations $\hat \zeta^\ha$, as a gauge potential under $U(1)$ 
transformations $\hat \chi$ and with the pullback of a shift under the NS-NS gauge 
transformations of the Kalb-Ramond field $\delta B_\hmn = 2 \part_{[\hmu}\Sigma_{\hnu]}$:
\be
\delta \hV_\ha = \hzeta^\hb \partial_\hb \hV_\ha 
             \ + \ \partial_\ha \hzeta^\hb \hV_\hb
             \ + \ \partial_\ha \hchi
             \ - \ \Sigma_\hmu \partial_\ha Y^\hmu. 
\label{deltahV}
\ee
It is the last term that makes the quantity $\hat \cF_{\ha\hb} = \hF_{\ha\hb} 
+ B_{\hmn} \partial_\ha Y^\hmu\ \partial_\hb Y^\hnu$ in the D-brane effective action 
invariant under NS-NS gauge transformations.

It is easy to show, T-dualising the $a$-components of the above equation, that the variations 
of the BI vector $V_a$ of the D$(p-1)$-brane are of the same form
\be
\delta V_a = \zeta^b \partial_b V_a 
             \ + \ \partial_a \zeta^b V_b
             \ + \ \partial_a \chi
             \ - \ \Sigma_\mu \partial_a X^\mu, 
\ee
given that the gauge parameters transform under T-duality as
\bea
\begin{array}{lll}
\hzeta^a \longrightarrow \zeta^a,  \hsp{2cm}  
&  \Sigma_\mu \longrightarrow \Sigma_\mu, \hsp{1.8cm} 
& \hchi \longrightarrow \chi \ -\ \Sigma_\ux X^\ux,\\[.2cm]
\hzeta^x \longrightarrow \Sigma_\ux,
  & \Sigma_\ux \longrightarrow \xi^\ux,
\end{array}
\label{Tdualparameters}
\eea
where we denote by $\xi^\hmu$ the parameter of general coordinate transformations 
$\delta x^\hmu = - \xi^\hmu$ in the target space. The interchange between parameters of 
coordinate transformations $\hzeta$ and $\xi$, and NS-NS gauge transformations $\Sigma$ is 
a consequence 
of the interchange of components of the metric and components of the Kalb-Ramond field by
T-duality.

With the same duality rules we can also derive the variations of the embedding scalars. From 
the variation of $\hV_x$, we see that $X^\ux$ transforms as a scalar under worldvolume
coordinate transformations, but as a coordinate under target space diffeomorphisms: 
\be
\delta X^\ux \ = \ \zeta^b \partial_b X^\ux  \  - \ \xi^\ux. 
\ee  
The form of the transformation rule for $X^\ux$ suggests that the variation of the embedding 
scalars in general are given by:
\be
\delta X^\hmu \ = \ \zeta^b \partial_b X^\hmu  \  - \ \xi^\hmu. 
\label{varX}
\ee  
Indeed, assigning the same status to the $Y^\hmu$'s as to the $X^\hmu$'s by supposing that 
their variations are given by   
\be
\delta Y^\hmu \ = \ \hzeta^\hb \partial_\hb Y^\hmu  \  - \ \xi^\hmu .
\ee
Applying T-duality to $Y^i$ we find exactly the equation (\ref{varX}).

Note that where the interchange of gauge transformations and diffeomorphisms in the Abelian 
case is an obvious consequence of the T-duality rules, the same property might become 
especially interesting in the non-Abelian case, where the concept of general coordinate 
transformations is much more difficult to understand. We will come to this point later on in 
this section.

\vspace{.3cm}
Much of the above story holds also for the non-Abelian case. The T-duality rule for the 
worldvolume fields (\ref{TdualWV}) and the gauge parameters (\ref{Tdualparameters}) are 
the same as for the Abelian case, as they match the degrees of freedom in both actions.  
The main difference however lies in the duality rules for the $U(N)$ covariant pullbacks 
$\hD_\ha Y^\hmu = \partial_\ha Y^\hmu + i [\hV_\ha, Y^\hmu]$ and the Yang-Mills field 
strengths $\hat F_{\ha\hb} = 2 \part_{[\hat a} V_{\hat b]} + i[\hV_\ha, \hV_\hb]$. In the 
non-Abelian case, these rules are given by
\bea
&&
\hD_a Y^i \longrightarrow D_a X^i, \hsp{1cm}
\hD_x Y^i \longrightarrow \ i[X^\ux, X^i], \hsp{1cm}
\part_{\hat a} Y^\ux = \delta^\ux_{\hat a}, \nn
&&
\hsp{1.5cm}
\hat F_{ax} \longrightarrow D_a X^\ux, \hsp{2cm}
\hat F_{ab}  \longrightarrow F_{ab}.
\label{TdualfieldstrNA}
\eea
As was first realised in \cite{Myers}, due to the non-trivial commutator term in the 
covariant derivative in the pullback, there is a non-trivial contribution from the 
dualisation of 
\be
\hD_x Y^i =  i [V_ x, Y^i] \ \longrightarrow \ i [X^\ux, X^i]. 
\ee
These extra contributions give rise to the so-called dielectric terms in the non-Abelian 
D-brane actions, enabling a D$p$-brane to couple to R-R field of rank $n>p+1$.

It is not difficult to see that a similar effect happens when deriving the duality rules 
for the gauge parameters (\ref{Tdualparameters}). The non-Abelian generalisation of the 
variations (\ref{deltahV}) of the BI vector $\hV_\ha$ are given by
\be
\delta \hV_\ha = \hzeta^\hb \partial_\hb \hV_\ha 
             \ + \ \partial_\ha \hzeta^\hb \hV_\hb
             \ + \ \hD_\ha \hchi
             \ - \ \Sigma_\hmu \hD_\ha Y^\hmu. 
\label{deltahVNA}
\ee  
Note that $\hV_\ha$ still transforms as a vector under the general coordinate transformations 
of the worldvolume coordinates (which remain Abelian), but has been promoted to the Yang-Mills 
vector of the $U(N)$ gauge group. Consistently with the pullbacks of the background fields 
in the actions we have replaced the pullback of the NS-NS gauge parameter $\Sigma_\hmu$ by a 
covariant pullback.\footnote{This expression for the NS-NS variation of $V$ was motivated in 
\cite{AGJL} by demanding $\cF= F + P[B]$ to be invariant, as the variation of $P[B]$ was 
believed to be $\delta P[B] = 2DP[\Sigma]$. Here, however, we do not wish to use this argument, 
since it will receive considerable corrections, as we will see shortly. A more correct 
argument is that this term in the variation is necessary to obtain the same form after T-duality.} 
A straightforward calculation, dualising the $\hV_a$-components of (\ref{deltahVNA}), shows 
that the variation of the BI vector of the D$(p-1)$-brane is of the same form:
\be
\delta V_a \ = \ \zeta^b \partial_b V_a 
             \ + \ \partial_a \zeta^b V_b
             \ + \ D_a \chi
             \ - \ \Sigma_\hmu D_a X^\hmu, 
\ee 
where again we used the T-duality rules (\ref{Tdualparameters}) for the gauge parameters. 

In the same way, T-dualising the $\hV_x$ component of (\ref{deltahVNA}), we obtain 
the variations of $X^\ux$:
\be
\delta X^\ux \ = \ \zeta^b \partial_b X^\ux \ - \ \xi^\ux  
               \ + \ i [X^\ux, \chi] \ - \ i \Sigma_\hrho [X^\ux, X^\hrho], 
\ee
which suggests the following variation for the general embedding scalars $X^\hmu$:
\be
\delta X^\hmu \ = \ \zeta^b \partial_b X^\hmu \ - \ \xi^\hmu  
               \ + \ i [X^\hmu, \chi] \ + \ i \Sigma_\hrho [X^\hrho, X^\hmu]. 
\label{varXNA}
\ee
Indeed, analogously as in the Abelian case, supposing that the variation of $Y^\hmu$ is of 
the form (\ref{varXNA})   
\be
\delta Y^\hmu \ = \ \hzeta^\hb \partial_\hb Y^\hmu  \  - \ \xi^\hmu 
      \ + \ i [Y^\hmu, \chi] \ + \ i \Sigma_\hrho [Y^\hrho, Y^\hmu] 
\ee
and applying T-duality to $Y^i$, we find exactly the equation (\ref{varXNA}). 

Note that the variation of the embedding scalars is considerably more complicated than in 
the Abelian case. The first and the third term state that the $X$'s (or $Y$'s) behave as 
scalars under worldvolume coordinate transformations and sit in the adjoint of the $U(N)$ 
gauge group, as one expects. 

The second term of (\ref{varXNA})  says that the embedding scalars are in fact the 
coordinates in the target space and transform as such under target space diffeomorphisms. 
Note that we have not made any assumption on how non-commutative diffeomorphisms should be 
performed and that the above form of the variation is induced by T-duality. It is remarkable 
that, in spite of the difficulty of how to implement non-commutative general coordinate 
transformations, in our case the variation of the embedding scalars does not receive any 
non-Abelian corrections, in contrast to the proposals made in for example \cite{DSW} or 
\cite{CK}. Note however that the algebra under which the coordinates transform is different 
for \cite{CK}. There the coordinates fulfill the canonical commutation relations 
$[X^\hmu, X^\hnu]=i\theta^{\hmn}$, where as in our case they are scalars in the adjoint 
representation of $U(N)$. It is known that the way of how to generalise non-commutative 
diffeomorphisms depends on the algebra satisfied by the coordinates.

Finally, the last term of (\ref{varXNA}) is a variation due to the NS-NS gauge transformation 
of the Kalb-Ramond field $\delta B_\hmn = 2\partial_{[\hmu}\Sigma_{\hnu]}$. This term is 
entirely non-Abelian and it arises exactly the same way as the dielectric couplings in 
\cite{Myers}, namely due to the non-trivial commutator term in the pullback 
$\Sigma_\hmu D_x Y^\hmu$ in (\ref{deltahVNA}). The existence of this term was already 
suggested in \cite{ere}, though its interpretation is less clear. In \cite{Hassan, DSW} it 
was observed that from the string theory point of view it is not clear whether transformations 
of the embedding scalars of the form $\delta X \sim [X, X]$ are in fact geometrical, i.e. 
whether they are really coordinate transformations. Our derivation, made above, seems to 
indicate that this is indeed not the case and that (at least part of) these transformations 
should be seen rather as NS-NS gauge transformations acting on the embedding scalars, given 
that they transform with the same parameter as the Kalb-Ramond field $B_\hmn$. 

It has also been suggested in \cite{ere} that these transformations could take care of the 
NS-NS gauge transformations of the terms of the form $[X^\hmu, X^\hnu] B_\hmn$ and 
$[X^\hmu, X^\hrho] B_\hmn$ in the Chern-Simons term of the non-Abelian Dp-branes. As 
explained in the introduction, these are precisely the terms that do not combine into 
the (apparently) gauge invariant quantity $\cF = F + P[B]$ and therefore seem to challenge 
the NS-NS gauge invariance of the non-Abelian Chern-Simons action. Note however that the 
situation is much more complex now, since together with the embedding scalars, also the 
pullbacks, the dielectric couplings and even the bulk fields will transform under the 
NS-NS transformations. We will show in the following sections that the action is in fact 
gauge invariant if one take into 
account the variations of the all these constituents.

\sect{NS-NS transformations of the fields}

Before we will show the NS-NS gauge invariance of non-Abelian Chern-Simons action, it is 
convenient to have a closer look on how the different constituents of the action transform
under the $\Sigma_\hmu$-transformations. It will be clear that, once the embedding scalars 
transform under the variation, all fields and quantities that depend on them will start to 
transform as well. 

It is straightforward to show that under the NS-NS gauge transformation
\be
\delta B_\mn = 2 \partial_{[\mu}\Sigma_{\nu]}, \hsp{1.5cm}
\delta V_a = -\Sigma_\mu D_a X^\mu, \hsp{1.5cm}
\delta X^\mu = i \Sigma_\rho [X^\rho, X^\mu], 
\label{Sigmavar}
\ee
the BI field strength, the $U(N)$ covariant pullback and the commutator transform as follows:
\bea
 \delta F_{ab} &=& - 2D_ {[a} ( \Sigma_\mu D_{b]}X^\mu ), \nn
\delta(D_a X^\mu) &=& i D_a (\Sigma_\rho [X^\rho, X^\mu]) 
                       \ - \ i [\Sigma_\rho D_a X^\rho, X^\mu], \label{varF} \\ [.2cm]
\delta [X^\mu, X^\nu] &=& i \Sigma_\rho [X^\rho, [X^\mu, X^\nu]] 
                       \ - \ i [\Sigma_\rho, X^\mu][X^\rho, X^\nu]
                       \ + \ i [\Sigma_\rho, X^\nu][X^\rho, X^\mu].\nonumber
\eea 
As a target space field $\Phi$ depends on the non-Abelian coordinates via the non-Abelian 
Taylor expansion \cite{Douglas2, GM}
\be
\Phi(X^\lambda) = \sum_n \frac{1}{n!} \partial_{\mu_1} ... \partial_{\mu_n} 
               \Phi(x^\lambda)|_{x^\lambda=0} X^{\mu_1}...X^{\mu_n},
\label{NATaylor}
\ee
the variation of this field under the transformations (\ref{Sigmavar}) is given by
\be
\delta \Phi = i \Sigma_\rho[X^\rho, \Phi].
\label{varPhi}
\ee
Note that the Kalb-Ramond field has an extra term due to its own gauge variation. Its full 
variation is therefore given by
\be
\delta B_\mn = 2\partial_{[\mu}\Sigma_{\nu]}\ + \ i \Sigma_\rho[X^\rho, B_\mn].
\label{varB}
\ee
Using the properties of the non-Abelian Taylor expansion (\ref{NATaylor}), one can derive 
the following useful identities for general background fields
\bea
D_a \Phi(X) = \partial_\mu \Phi (X) \ D_a X^\mu, \hsp{2cm}
[\Phi (X), X^\mu] = \partial_\rho \Phi (X)\ [X^\rho, X^\mu].
\label{properties}
\eea 
The Taylor expansion (\ref{NATaylor}) requires a symmetrised trace prescription \cite{Tseytlin}, 
so the identities (\ref{properties}) are only valid when performed inside the action. 

Taking in account the restriction of the symmetrised trace, one can use the identities 
(\ref{properties}) to rewrite the variations (\ref{Sigmavar}), (\ref{varF}) and (\ref{varPhi}), 
yielding
\bea
\delta V_a &=& -\Sigma_\mu D_a X^\mu \nn
\delta X^\mu &=& i[X^\rho, \Sigma_\rho X^\mu] 
       \ + \ i \partial_{[\sigma}\Sigma_{\rho]}[X^\sigma, X^\rho] X^\mu,
 \nn
\delta F_{ab} &=& i[X^\rho, \Sigma_\rho F_{ab}] 
       \ + \ i \partial_{[\sigma}\Sigma_{\rho]}[X^\sigma, X^\rho] F_{ab}
       \ - \ 2 \partial_{[\sigma}\Sigma_{\rho]} D_{[a}X^\sigma D_{b]}X^\rho, 
\label{varF2} \\ [.2cm]
\delta D_a X^\mu &=& i[X^\rho, \Sigma_\rho D_{a} X^\mu] 
       \ + \ i \partial_{[\sigma}\Sigma_{\rho]}[X^\sigma, X^\rho] D_{a} X^\mu
       \ + \ 2 i  \partial_{[\sigma}\Sigma_{\rho]} D_{a}X^\sigma [X^\rho, X^\mu], \nn
\delta [X^\mu, X^\nu] &=& i[X^\rho, \Sigma_\rho [X^\mu, X^\nu]] 
       \ + \ i \partial_{[\sigma}\Sigma_{\rho]}[X^\sigma, X^\rho] [X^\mu, X^\nu]
       \ - \ 2 i \partial_{[\sigma}\Sigma_{\rho]} [X^\sigma, X^\mu][X^\rho, X^\nu], \nn
\delta\Phi &=& i[X^\rho, \Sigma_\rho \Phi] 
       \ + \ i \partial_{[\sigma}\Sigma_{\rho]}[X^\sigma, X^\rho] \Phi.\nonumber
\eea  
Note that, except for the variation of $V_a$, all variations have the same structure:
\be
\delta Z = i[X^\rho, \Sigma_\rho Z] 
       \ + \ i \partial_{[\sigma}\Sigma_{\rho]}[X^\sigma, X^\rho] Z
       \ + \ \mbox{possible correction terms}, 
\label{varZ}
\ee
Furthermore the second term in the variation is proportional to 
$(\incl_X \incl_X \partial\Sigma)$, which has the same form as what one naively would say is 
the variation of $(\incl_X \incl_X B)$, one of the terms that do not combine into an $\cF$. 
On the other hand, the correction terms, present in the variation of $F$, $DX$ and $[X, X]$
are proportional to different contractions of $\partial \Sigma$ with the embedding scalars. 
We will see in the next section that the variations have precisely the correct structure to 
ensure the  gauge invariance of the action.

Finally it is worthwhile to notice that the general structure of the transformations given 
in (\ref{varZ}) holds, even for the case where $Z$ is a composite object. It is not difficult 
to show that in that case we have that
\bea
\delta (Z_1Z_2) &=& i[X^\rho, \Sigma_\rho (Z_1Z_2)] 
       \ + \ i \partial_{[\sigma}\Sigma_{\rho]}[X^\sigma, X^\rho] (Z_1Z_2)\nn
       && \hsp{.5cm} + \ \mbox{possible correction terms due to $Z_1$} \label{varZZ}\\ [.2cm]
       && \hsp{.5cm} + \ \mbox{possible correction terms due to $Z_2$} , \nonumber
\eea

To summarise, we give in shorthand notation the variations (\ref{varF2}) of the objects 
that appear in the action:
\bea
\delta V &=& -P[\Sigma], \nn
\delta X^. &=& i[X, \Sigma X^.] -2 i(\incl_X\incl_X \partial\Sigma) X^. , \nn
\delta F &=& i[X, \Sigma F] -2 i(\incl_X\incl_X \partial\Sigma) F
              - 2 P[\partial\Sigma] , \label{varSigma3} \\ [.2cm]
\delta DX^. &=& i[X, \Sigma DX^.] -2 i(\incl_X\incl_X \partial\Sigma) DX^. 
              - 4 i P[(\incl_X \partial\Sigma) \incl_X . ], \nn
\delta (\incl_X \incl_X. ) &=&i[X, \Sigma (\incl_X \incl_X . )] 
              -2i (\incl_X\incl_X \partial\Sigma) (\incl_X \incl_X .) 
              + 4 i  P[(\incl_X \incl_Y \partial\Sigma) \incl_Y \incl_X . ], \nn
\delta \Phi &=& i[X, \Sigma \Phi] -2i (\incl_X\incl_X \partial\Sigma) \Phi.\nonumber
\eea
The convention for the last term in the variations of $DX$ and $[X, X]$ should be clear from 
(\ref{varF2}), but for the sake of clearness give our convention explictly: 
\bea
&&P[(\incl_X \partial\Sigma) \incl_X . ] =\half [X^\sigma, X^\mu] \partial_{[\sigma} \Sigma_{\rho]}, 
\nn
&&(\incl_X \incl_Y \part \Sigma) (\incl_Y \incl_X C_2) 
=  \tfrac{1}{4} [X^\nu, X^\sigma][X^\rho, X^\mu]\partial_{[\sigma} \Sigma_{\rho]} C_\mn.
\eea 
We recall that strictly speaking the above variation rules are only valid when used inside a 
symmetrised trace.

\sect{Invariance of the D6-brane action}

In this section we will prove the gauge invariance of the non-Abelian Chern-Simons action of 
the  D6-brane, as it is the simplest non-trivial case. (For a general proof of the invariance 
of the D$p$-brane action we refer to Appendix A.)  It is however instructive to first consider 
the variation of a typical term from the action.

Given the transformation rules (\ref{varSigma3}), it is not difficult to see that the 
variation of a typical term 
\be
\cL \ = \ \Str \Bigl\{ P\Bigl[ (\incl_X \incl_X)^r \Bigr(C_p B^k\Bigr)\Bigr] F^m \Bigr\}  
\label{typterm}
\ee
in the Chern-Simons action will transform as follows:
\bea
\delta \cL &=&  \Str \Bigl\{ i [X^\rho, \Sigma_\rho \cL] 
\ - \ 2i P\Bigl[ (\incl_X \incl_X\partial\Sigma)(\incl_X \incl_X)^r \Bigr(C_p B^k\Bigr)\Bigr] F^m 
\nn
&&\hspace*{1cm}
+ \ 4 i r \ P\Bigl[ (\incl_X \incl_Y \partial\Sigma)
             (\incl_Y \incl_X)(\incl_X \incl_X)^{r-1} \Bigr(C_p B^k\Bigr)\Bigr] F^m \nn
&&\hspace*{1cm}
+ \ 2 k  \ P\Bigl[ (\incl_X \incl_X)^r \Bigr(C_p B^{k-1} \partial\Sigma \Bigr)\Bigr] F^m 
\label{varterm} \\ [.2cm]
&&\hspace*{1cm}
- \ 4 i (p-2r+2k) \ P \Bigl[ (\incl_X \partial\Sigma) 
                    \incl_X (\incl_X \incl_X)^r \Bigr(C_p B^k\Bigr)\Bigr] F^m \nn
&&\hspace*{1cm}
- \ 2 m \ P \Bigl[(\incl_X \incl_X)^r \Bigr(C_p B^k\Bigr)\partial \Sigma \Bigr] F^{m-1} \Bigr\}, 
\nonumber  
\eea
where the first two terms come from the  general structure of the variations, the third term 
from the correction term in the variation of $[X, X]$, the fourth one from the variation of 
$B_\mn$, the fifth from the correction term in the variation of the $DX$'s and the last term 
from the correction term in the variation of $F$.

Making use of the fact that the different $(\incl_X \incl_X)$ contractions are distributed over 
the antisymmetrised combination of background fields $(C_p B^k\partial\Sigma)$ as
\bea
&& (p + 2k +2)(p + 2k + 1) \ (\incl_X \incl_X)^{r+ 1}\Bigl(C_p B^k\partial\Sigma \Bigr)=  \nn 
&&   \hsp{1.5cm}
   = \ 2 (r+1) \ (\incl_X \incl_X)^{r}\Bigl(C_p B^k \Bigr) 
             \Bigl(\incl_X \incl_X \partial\Sigma \Bigr)\nn
&&\hsp{1.5cm}
 \ - \ 4 r (r+1)\  (\incl_X \incl_Y)(\incl_X \incl_X)^{r-1}\Bigl(C_p B^k \Bigr) 
             \Bigl(\incl_Y \incl_X \partial\Sigma \Bigr) \\
&&   \hsp{1.5cm}
 \ + \ 4 (r+1)(p-2r+2k)\  \Bigl(\incl_X  \partial\Sigma \Bigr)
          \incl_X (\incl_X \incl_X)^{r}\Bigl(C_p B^k \Bigr)
        \nn
&&   \hsp{1.5cm} 
 \ +\  (p-2r+2k)(p-2r+2k-1) \ (\incl_X \incl_X)^{r+ 1}\Bigl(C_p B^k\Bigr)\partial\Sigma, 
\nonumber
\eea
we can combine the second, the third and the fifth term of (\ref{varterm}) and rewrite the 
expression as 
\bea
\delta \cL &=& \Str\Bigr\{
 \ 2 k  \ P\Bigl[ (\incl_X \incl_X)^r \Bigr(C_p B^{k-1} \partial\Sigma \Bigr)\Bigr] F^m 
\ - \ 2 m \ P \Bigl[(\incl_X \incl_X)^r \Bigr(C_p B^k\Bigr)\partial \Sigma \Bigr] F^{m-1}
\nn
&& \hsp{1.5cm}\ -\ i\ \tfrac{(p+2k+2)(p+2k+1)}{(r+1)} 
      \ P\Bigl[(\incl_X \incl_X)^{r+ 1}\Bigl(C_p B^k\partial\Sigma \Bigr)\Bigr] F^m 
\label{varterm2}
\\ [.2cm]
&& \hsp{1.5cm} \ + \ i \ \tfrac{(p-2r+2k)(p-2r+2k-1)}{(r+1)}
       \ P\Bigl[(\incl_X \incl_X)^{r+ 1}\Bigl(C_p B^k \Bigr)\partial\Sigma\Bigr] F^m  \Bigr\}. 
\nonumber
\eea 
Note that we have omitted the first term of (\ref{varterm}), since the trace of a single 
commutator of $U(N)$ matrices is zero. We then see that the variation of a typical term
(\ref{typterm}) can be written as contractions of commutator couplings with the 
antisymmetrised combination $(C_p B^k \partial \Sigma)$ or contractions of $(C_p B^k)$ 
antisymmetrised with $\partial \Sigma$. 

The point now is that the various contributions in the variation of terms with the same 
value for $p$, but with different values for $r$, $k$ and $m$ will cancel the terms in  
(\ref{varterm2}). We will show that the coefficients of each term are such that the 
respective terms precisely cancel. 


The Chern-Simons part of the non-Abelian D6-brane action can be written as \cite{Myers}
\be
\cL_{D6} = \Str\Bigl\{ \sum_{r=0}^{1} \sum_{n=0}^{3+r} \sum_{k=0^*}^n \alpha_{rnk}
      P \Bigl[ (\incl_X\incl_X)^r \Bigl(C_{7+2r-2n}B^k\Bigr)\Bigr]F^{n-k}\Bigr\}, 
\ee
where the coefficients $\alpha_{rnk}$ are given by
\be
\alpha_{rnk}= \frac{(-1)^{n+r} \ i^r \ 7! (7+2r-2n+2k)!}
                   {2^n (n-k)!\ k! \ r! \ (7-2n+2k)!(7+2r-2n)!}\ .
\ee
The star in the summation over $k$ indicates that for the terms with $r=1$ and $n=4$ the  
lowest possible value is actually $k=1$ rather then $k=0$.

Making use of (\ref{varterm2}), the variation of $\cL_{D6}$ yields
\bea
\delta \cL_{D6} &=&  \sum_{r,n,k} \ \alpha_{rnk} \  \Str\Bigl\{ \ 
      2k\   P\Bigl[ (\incl_X \incl_X)^r \Bigl( C_{7+2r-2n}B^{k-1}\partial\Sigma\Bigr)\Bigr]F^{n-k}
\nn
&& 
- \ 2(n-k)\ P\Bigl[ (\incl_X \incl_X)^r \Bigl( C_{7+2r-2n}B^{k}\Bigr)\partial\Sigma\Bigr]F^{n-k-1}
\\ [.2cm]
&& 
- \ i \ \tfrac{(9+2r-2n+2k)(8+2r-2n+2k)}{(r+1)}  
    \  P\Bigl[ (\incl_X \incl_X)^{r+1} \Bigl( C_{7+2r-2n}B^{k}\partial\Sigma\Bigr)\Bigr]F^{n-k}
\nn
&& 
+ \ i \ \tfrac{(7-2n+2k)(6-2n+2k)}{(r+1)}
         \ P\Bigl[ (\incl_X \incl_X)^{r+1} \Bigl( C_{7+2r-2n}B^{k}\Bigr)\partial\Sigma\Bigr]F^{n-k}
\Bigr\}, \nonumber
\eea
where we omitted the single commutator term, as it vanishes upon taking the trace.
The $r=1$ contributions in the last two term are zero, since on the one hand we have four 
embedding scalars contracted with an antisymmetrised combination of background fields, while 
on the other hand the D6-brane only has three transverse (and hence non-Abelian) scalars.

Writing out explicitly the remaining terms, we find
\bea
\delta \cL_{D6} &=&   \Str\Bigl\{ \ 
\sum_{n=1}^{3} \sum_{k=1}^{n} \tfrac{(-)^n 7!}{2^{n-1} (n-k)! (k-1)! (7-2n)! }
     \ P\Bigl[ C_{7-2n} B^{k-1} \partial\Sigma\Bigr] F^{n-k} \nn
&& \hsp{.7cm}
  + \sum_{n=1}^{4} \sum_{k=1}^{n} 
        \tfrac{(-)^{n+1} i  \ 7! \ (9-2n+2k)!}{2^{n-1}(n-k)! (k-1)!(7-2n+2k)!(9-2n)!}
     \ P\Bigl[ (\incl_X\incl_X)\Bigl(C_{9-2n} B^{k-1} \partial\Sigma\Bigr) \Bigr] F^{n-k} \nn
&& \hsp{.7cm}
  + \sum_{n=1}^{3} \sum_{k=0}^{n-1} 
        \tfrac{(-)^{n+1}  7!}{2^{n-1}(n-k-1)! k!(7-2n)!}
     \ P\Bigl[ C_{7-2n} B^{k} \partial\Sigma\Bigr] F^{n-k-1} \\ [.2cm]
&& \hsp{.7cm}
 + \sum_{n=1}^{4} \sum_{k=0^*}^{n-1} 
        \tfrac{(-)^{n+2} i \ 7! \ (9-2n+2k)!}{2^{n-1}(n-k-1)! k!(7-2n+2k)!(9-2n)!}
     \ P\Bigl[ (\incl_X\incl_X)\Bigl(C_{9-2n} B^{k} \Bigr)\partial\Sigma \Bigr] F^{n-k-1} \nn
&& \hsp{.7cm}
  + \sum_{n=0}^{3} \sum_{k=0}^{n} 
        \tfrac{(-)^{n+1} i  \ 7! \ (9-2n+2k)!}{2^{n}(n-k)! k!(7-2n+2k)!(7-2n)!}
     \ P\Bigl[ (\incl_X\incl_X)\Bigl(C_{7-2n} B^{k} \partial\Sigma\Bigr) \Bigr] F^{n-k} \nn
&& \hsp{.7cm}
 + \sum_{n=0}^{3} \sum_{k=0^*}^{n} 
        \tfrac{(-)^{n} i \ 7! \ (7-2n+2k)!}{2^{n}(n-k)! k!(5-2n+2k)!(7-2n)!}
     \ P\Bigl[ (\incl_X\incl_X)\Bigl(C_{7-2n} B^{k} \Bigr)\partial\Sigma \Bigr] F^{n-k}
\ \Bigl\} \ = \ 0, \nonumber
\eea 
as the first term cancels the third after relabeling $k-1 = l$, the second term cancels the 
fifth after relabeling $n-1=m$ and $k-1=l$ and the the fourth term cancels the last one after 
relabeling $n-1= m$.

In summary, the variation under the NS-NS variations of the D6-brane vanishes exactly, as 
it can be written  as a series of terms which cancel each other exactly, plus a series of terms 
that are identically zero, either because they form a single commutator, or because they 
contain more dielectric couplings than the transverse directions actually permit.

The general result for the invariance of the D$p$-brane is completely analogous, though it 
involves a few more terms which are not present in this simple case. We will not give the 
full proof here and refer the interested reader to appendix A.

\sect{Discussion}

We have seen that the requirement that the worldvolume fields in the D$p$-brane action 
play the same role as in the D$(p-1)$-brane, as T-duality suggests, induced non-Abelian 
variations for the embedding scalars $\delta X^\mu \sim \Sigma_\rho [X^\rho, X^\mu]$. 
Given that the parameter $\Sigma_\mu$ is the same as the parameter of the gauge transformations 
of the Kalb-Ramond field $B_\mn$, these variations can be interpreted as NS-NS gauge 
transformations of the embedding scalars.

The fact that the embedding scalars transform under gauge transformations implies that 
almost all of the components of the action transform as well in a non-trivial way: the 
pullbacks and the dielectric couplings, as they are combinations of the $X$'s, but also 
the background fields, as they are functions of the embedding scalars, even while in 
supergravity they are inert under NS-NS transformations.

The immediate question then arises whether the non-Abelian D-brane action as given in 
\cite{Myers} is invariant under these NS-NS transformations. In particular: are 
these variations able to cancel the variations of the terms of the form $(\incl_X\incl_X)B$ 
and $\incl_X B$ which were ignored in the study of the gauge invariance made in \cite{AGJL}?

We showed that this is indeed the case: the variations of the different components of a 
typical term $P[(\incl_X\incl_X)^r (C_p B^k)] F^m$ (pullbacks, dielectric couplings, BI 
field strength, and background fields) are such that they combine with the variation of 
other terms with different values for $r$, $k$ and $m$, and either cancel each other or 
form terms that vanish identically.  The remarkable feature therefore is that the gauge 
invariant combinations are not, as in the case of the R-R gauge invariance, blocks with 
the same number $r$ of dielectric couplings, but  blocks with the same rank $p$ of the R-R 
field. Furthermore it is remarkable that 
in spite of the non-trivial way each field transforms, the action is invariant in its 
present form. Therefore the conjecture of \cite{AGJL}, predicting the existence of extra 
terms coming from a proper covariant formulation of the action, turns out to be wrong, at 
least in so far as they are needed to obtain an action which is invariant under the NS-NS 
transformations.

It is worth noticing that the way the NS-NS transformations act on the background 
fields is quite different from how the R-R transformations. In \cite{AGJL} it was proposed 
that the variations of the pullback of a R-R field is the derivative of the pullback of the 
gauge parameter 
\be
\delta P[(\incl_X\incl_X)^r C_p] = D P[(\incl_X\incl_X)^r \Lambda_{p-1}].
\label{defRR}
\ee 
Actually, this was seen as the very definition of how to implement the R-R gauge transformation 
in the non-Abelian worldvolume, as it is the only way to assure the the variation is a total 
derivative, a scalar under $U(N)$ and reduces to the known Abelian case. In other words, the 
implementation of the transformation rules had to be changed in order to obtain an invariant 
action. However in the case of the NS-NS gauge invariance, the transformation rules of each 
of the components was derived through T-duality and the variation of a typical term is just 
the sum of the variations of its components. For example the difference can be clearly seen 
when comparing the defined variation of $P[C_2]$ and the derived variation of $P[B_2]$, 
which are supposed to be each other S-duals. Acting with the covariant derivative of 
(\ref{defRR}) inside the pullback, we have
\bea
\delta P[C_2] &=&  2 P[\partial \Lambda_1] \ - \  i \Lambda_1 [X, F], \nn
\delta P[B_2] &=& 2P[\partial\Sigma] \ - \ 2i P[(\incl_X\incl_X \partial\Sigma)B]
                              \ - \ 8i P[(\incl_X \partial\Sigma) (\incl_X B)] .
\label{compareCB}
\eea  
However one has to bear in mind that S-duality takes the action beyond the perturbative regime 
and that it is not clear how to incorporate this symmetry in a non-Abelian theory.
It is therefore not surprising that, in the actions considered here, fields that belong to 
different sectors transform in different ways under gauge transformations. In a fully 
non-perturbative description, both expressions (\ref{compareCB}) will most likely receive 
corrections, such that the symmetry is restored.

Finally it is not difficult to see that the corrections found in the NS-NS variation do
not induce corrections in the massive gauge transformations in $m$IIA theory, even though 
these have the same gauge parameter $\Sigma$. We can therefore conclude that the non-Abelian 
Chern-Simons action for D-branes as presented in \cite{Myers} and completed with the mass term
of \cite{GHT} is invariant under the complete R-R, NS-NS and massive gauge transformations
of the background fields.


\vsp{1cm}

\noindent
{\bf Acknowledgments}\\
\vspace{-.2cm}

\noindent
We wish to thank Yolanda Lozano and Rob Myers for the useful discussions.
The work of J.A. is done as Aspirant F.W.O. (Belgium). She is also partially supported 
by the European Commission FP6 program MRTN-CT-2004-005104, by the F.W.O.-Vlaanderen 
project G0193.00N and by the Belgian Federal Office for Scientific, Technical and 
Cultural Affairs through the Interuniversity Attraction Pole P5/27. 
The work of I.A.I. is supported by the M.E.C. (Spain) under the contract FPA2003-09298-C02-01. 
The work of B.J.~is done as part of the program ``Ram\'on y Cajal'' of the M.E.C and is 
partially supported by the M.E.C. under contract FIS 2004-06823. I.A.I. and B.J. are also 
partially supported by the Junta de Andaluc\'{\i}a under the contract FQM 101.

\appendix 
\renewcommand{\theequation}{\Alph{section}.\arabic{equation}}
\sect{Invariance of the general D$p$-brane action}

The proof of the invariance of the non-Abelian Chern-Simons action for D$p$-branes is 
analogous to and only slightly more complicated than the derivation given in section 4 for 
the D6-brane. We give the general result here for the sake of completeness. 

The Chern-Simons part of the general non-Abelian D$p$-brane action is given by
\be
\cL_{Dp} = \Str\Bigl\{ \sum_{r=0}^{[\frac{9-p}{2}]} \sum_{n=0}^{[\frac{p+2r+1}{2}]} \sum_{k=0^*}^n 
         \alpha_{rnk}
      P \Bigl[ (\incl_X\incl_X)^r \Bigl(C_{p+2r-2n+1} B^k\Bigr)\Bigr]F^{n-k}\Bigr\}, 
\ee
where the coefficients $\alpha_{rnk}$ are given by
\be
\alpha_{rnk}= \frac{(-1)^{n+r} \ i^r \ (p+1)! (p+2r-2n+2k+1)!}
                   {2^n (n-k)!\ k! \ r! \ (p-2n+2k+1)!(p+2r-2n+1)!}\ .
\ee
The square brackets in the upper summation limit refer to the integer part of the fraction, 
while the star in the summation over $k$ indicates that for the terms with $n> (p+2k+1)/2$ the  
lowest possible value is actually $k= (2n-p+1)/2$ rather then $k=0$.

The variation of the action is given by
\bea
\delta \cL_{Dp} &=& \Str\Bigl\{ 
\sum_{r=0}^{[\frac{9-p}{2}]} \sum_{n=1}^{[\frac{p+2r+1}{2}]} \sum_{k=1^*}^n 2k \ \alpha_{rnk} \ 
    P \Bigl[ (\incl_X\incl_X)^r \Bigl(C_{p+2r-2n+1} B^{k-1} \partial\Sigma \Bigr)\Bigr]F^{n-k} \nn
&& \hsp{.5cm}
+ \sum_{r=0}^{[\frac{9-p}{2}]} \sum_{n=1}^{[\frac{p+2r+1}{2}]} \sum_{k=0^*}^{n-1} 
       (-2)(n-k) \ \alpha_{rnk} \ 
    P \Bigl[ (\incl_X\incl_X)^r \Bigl(C_{p+2r-2n+1} B^{k}\Bigr) \partial\Sigma \Bigr]F^{n-k-1} \nn
&& \hsp{.5cm}
+ \sum_{r=0}^{[\frac{7-p}{2}]} \sum_{n=0}^{[\frac{p+2r+1}{2}]} \sum_{k=0^*}^{n}  
       (-i) \tfrac{(p+2r-2n+2k+3)(p+2r-2n+2k+2)}{(r+1)}\ \alpha_{rnk} \  \nn
&& \hsp{5cm}
    P \Bigl[ (\incl_X\incl_X)^{r+1} \Bigl(C_{p+2r-2n+1} B^{k} \partial\Sigma\Bigr) \Bigr]F^{n-k} \nn
&& \hsp{.5cm}
+ \sum_{r=0}^{[\frac{7-p}{2}]} \sum_{n=0}^{[\frac{p+2r+1}{2}]} \sum_{k=0^*}^{n}  
       i\ \tfrac{(p-2n+2k+1)(p-2n+2k)}{(r+1)}\ \alpha_{rnk} \  \nn
&& \hsp{5cm}
    P \Bigl[ (\incl_X\incl_X)^{r+1} \Bigl(C_{p+2r-2n+1} B^{k} \Bigr) \partial\Sigma \Bigr]F^{n-k}
\Bigr\}, 
\label{varaction}
\eea
where we have omitted the single commutator and the $r=[\frac{9-p}{2}]$ contributions of the 
last two terms as they carry a non-Abelian coupling of the form 
$(\incl_X\incl_X)^{[\frac{11-p}{2}]}$ while the D-brane only has $9-p$ transverse  directions. 

It is straightforward to see that the $r=0$ contribution of the first term cancels the $r=0$ 
contribution of the second term. Similarly, the $r\neq 0$ contributions of the first term 
cancel the third term, while the  $r\neq 0$ contributions of the second term cancel the fourth 
one.


\end{document}